\documentclass[10pt,twocolumn,conference, a4paper]{IEEEtran}

\usepackage{balance}

\usepackage{graphicx}
\usepackage{url}
\usepackage{amsmath,amssymb,bbm}
\usepackage{algpseudocode}

\def\QSet{{\cal Q}}
\def\RSet{{\cal R}}
\def\zZ{{\mathbb Z}}

\def\approx{\mbox{\footnotesize ap}}

\title{Optimal Quantization of TV White Space Regions for a
Broadcast Based Geolocation Database}
\author{\IEEEauthorblockN{Garima Maheshwari and Animesh Kumar}
\IEEEauthorblockA{Department of Electrical Engineering\\ Indian
Institute of Technology, Bombay\\ Mumbai, India -- 400076\\ Email:
garima.m@iitb.ac.in, animesh@ee.iitb.ac.in} }

\begin{document}

\maketitle

\begin{abstract}
In the current paradigm, TV white space databases communicate the
available channels over a reliable Internet connection to the
secondary devices. For places where an Internet connection is not
available, such as in developing countries, a broadcast based
geolocation database can be considered. This geolocation database will
broadcast the TV white space (or the primary services protection
regions) on rate-constrained digital channel. 

In this work, the quantization or digital representation of protection
regions is considered for rate-constrained broadcast geolocation
database. Protection regions should not be declared as white space
regions due to the quantization error. In this work, circular and
basis based approximations are presented for quantizing the protection
regions.  In circular approximation, quantization design algorithms
are presented to protect the primary from quantization error while
\textit{minimizing} the white space area declared as protected region.
An efficient quantizer design algorithm is presented in this case.
For basis based approximations, an efficient method to represent the
protection regions by an `envelope' is developed. By design this
envelope is a sparse approximation, i.e., it has lesser number of
non-zero coefficients in the basis when compared to the original
protection region.  The approximation methods presented in this work
are tested using three experimental data-sets.
\end{abstract}

\section{Introduction}
\label{sec:intro}

The wireless spectrum is a limited and valuable resource.  The demand for
spectrum is increasing due to the increase in the number of wireless devices and
this demand has led to research for efficient utilization techniques of the
spectrum. The usage of TV white space by unlicensed secondary users is an
example of efficient utilization of spectrum.  The spectrum licensing agencies,
Federal Communications Commission (FCC) in the United States and Office of
Communication (Ofcom) in the United Kingdom, have permitted access of TV white
space by an unlicensed secondary device \cite{FCC08260S2008,OfcomD2009}.

\begin{figure}[!htb]
\centering
\includegraphics[scale=1.0]{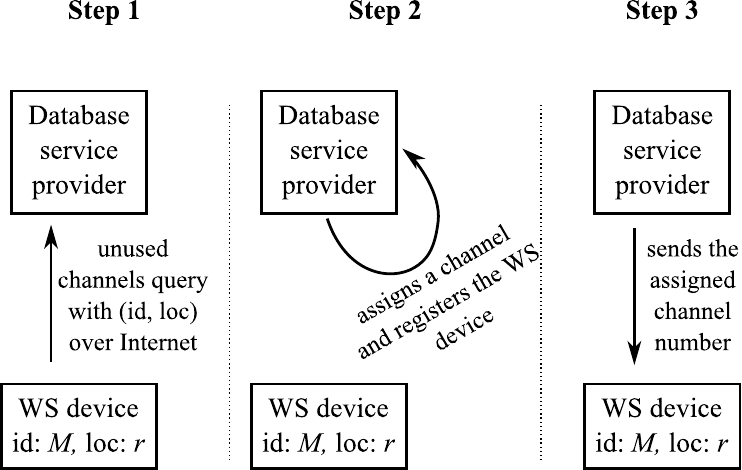}
\caption{\label{fig:dbprotocol} Three step explanation of database
operation is explained. A white space (WS) device queries the database
with geolocation information $r$ and device id $M$. The database
assigns an available channel and registers the device $M$.}
\end{figure}

According to the existing regulations of FCC and Ofcom, TV white space
can be accessed by a secondary or white space device (unlicensed user)
via TV white space (geolocation) database access.  A certified TV
white space database service is queried before operation by the
secondary device.  This query includes the location of secondary
device, and database registers the secondary (white space) device if
it is allocated a `white' TV channel. The TV transmitter protection
regions are calculated by the TV white space database service
providers to avoid harmful interference to the primary devices of the
licensed broadcasting services. The available TV white space changes
with time and space, and it is mandatory for the secondary device to
know the availability at the location and time of current operation.
The access of TV white space database takes place over the
Internet~\cite{FCC08260S2008}.  Fig.~\ref{fig:dbprotocol} depicts the
scheme for accessing TV white space database over the Internet.  By
design, TV white space database access is inaccessible for secondary
devices in areas where there is unreliable or no Internet connection.
The lack of internet connection is especially prevalent in many
developing or under-developed countries where internet services are
limited. In such areas, an \textit{alternate scheme} for communication
of the protection regions of TV transmitters are needed.

\begin{figure}[!htb]
\centering
\includegraphics[width=3.0in]{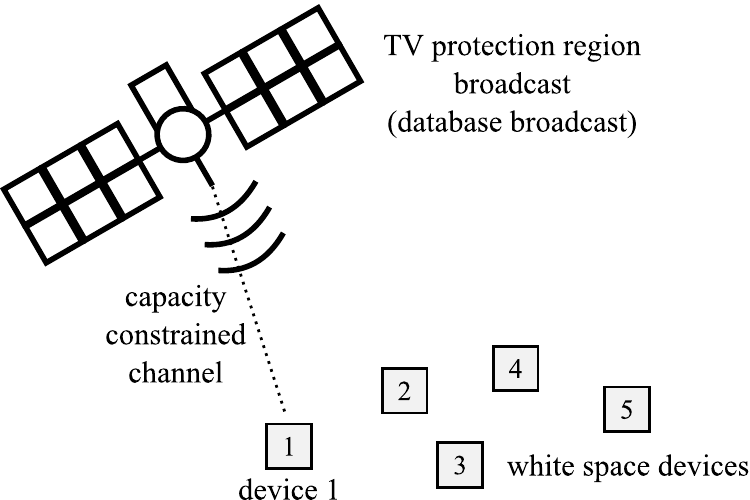}
\caption{\label{fig:schematic} A broadcast based geolocation database
is illustrated. It is assumed that white space devices can receive the
broadcast over a rate constrained channel. Thus, the database should
quantize the protection regions, while ensuring the protection of
primary from quantization error.}
\end{figure}

In this work, a \textit{broadcast based} database service for
communicating TV white space availability is proposed. A broadcast
based database, such as one using a satellite, can transmit the TV
transmitters' protection regions, or simply \textit{the protection
regions}. In this broadcast based database, the white space device
will not query and only receive the broadcast message from the
database. The proposed broadcast based database will communicate the
protection regions to all the white space devices in a region.  This
broadcast based approach is assumed to make use of a wireless or
digital channel, and the transfer of information to a secondary device
will be rate constrained (see Fig.~\ref{fig:schematic}).  Accordingly,
quantization of protection regions is of interest for a
broadcast-style TV white space database.  When the protection regions
are quantized, some TV white space area will be `lost' due to
quantization error. This quantizer has to be designed to
\textit{minimize} the TV white space area lost due to quantization. In
summary, protection region quantizer's \textit{design and performance}
are the key problems addressed in the current work.  \textit{To the
best of our knowledge, this is the first work of its kind.}

\begin{figure}[!htb]
\centering
\includegraphics[scale=0.4]{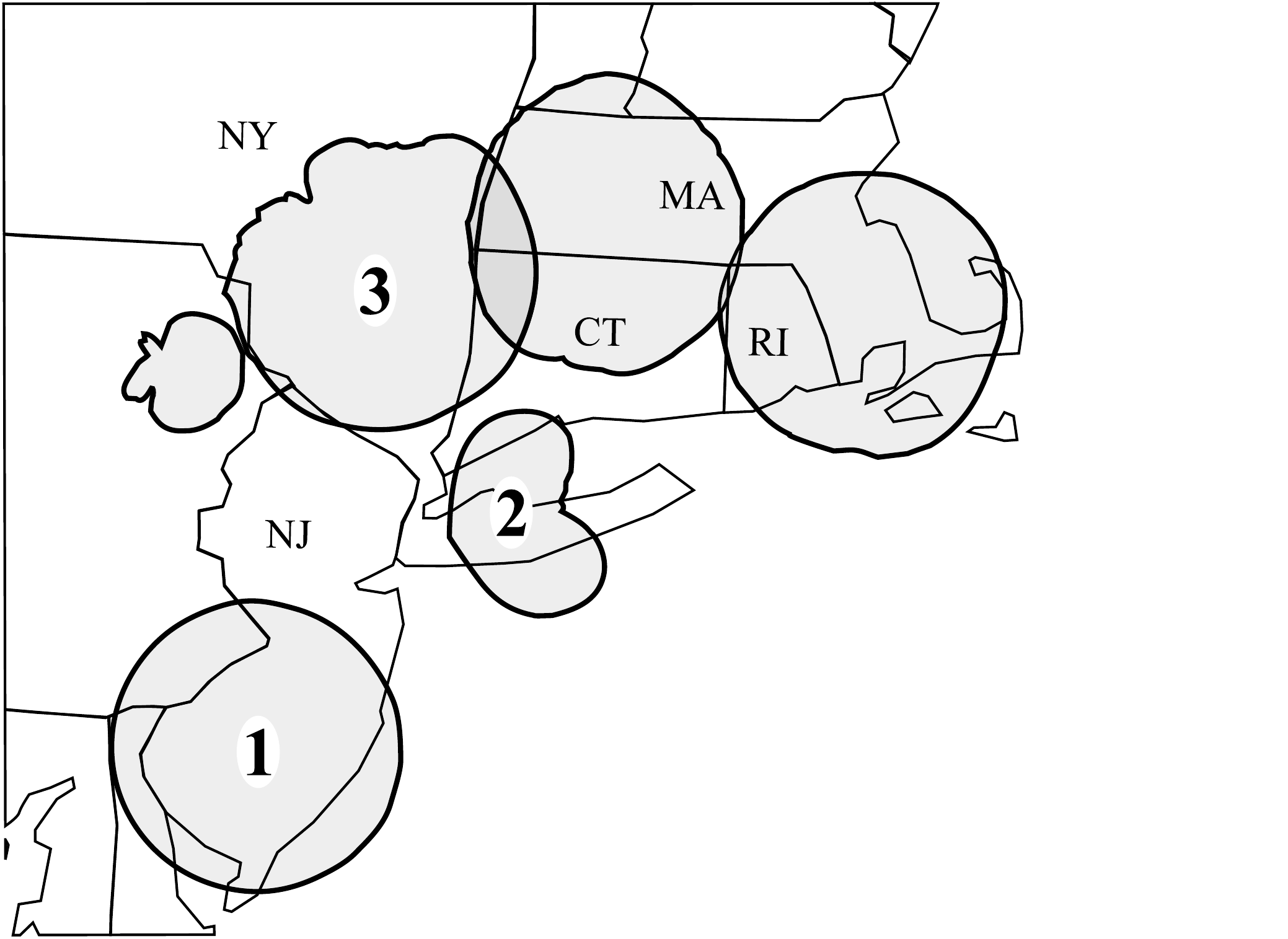}
\caption{\label{fig:coveragecontours} This graph is obtained from
the website of a certified TV white space service provider iconectiv
in the United States~\cite{iconectivWEB}. The contours are protection
regions for Channel~22 near New York. Protection region labeled
\textbf{1} is almost circular while protection regions labeled
\textbf{2} and \textbf{3} are non-circular in shape.}
\end{figure}

\begin{figure}[!htb]
\centering
\includegraphics[scale=1.0]{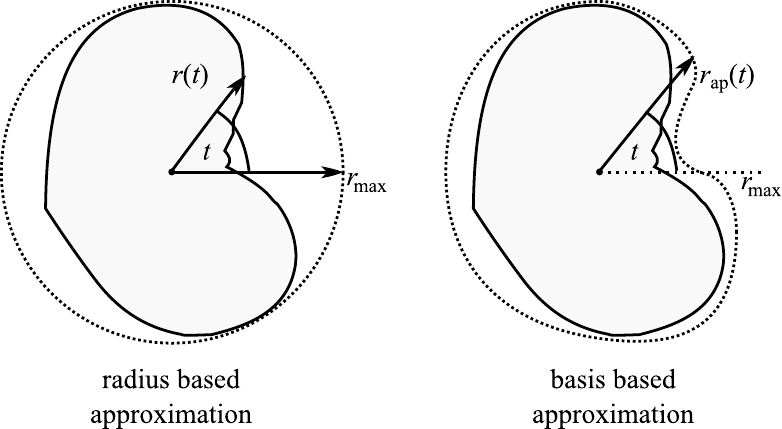}
\caption{\label{fig:naivesmart} The circular approximation and basis
based approximation approaches are illustrated. In the circular
approximation approach (left figure), 
the protection region is enlarged and made circular by 
inscribing it in a circle with maximum radius. In the basis based
approximation approach, an approximation $r_{\approx}(t)$ is obtained
such that the protection region is a subset of this approximation.}
\end{figure}

Examples of the protection regions are illustrated in
Fig.~\ref{fig:coveragecontours}, which are obtained from the iconectiv
website~\cite{iconectivWEB} for Channel~22 in the New York region. Protection
regions such as \textbf{1} are nearly circular while protection
regions such as \textbf{2} and \textbf{3} are non-circular (see
Fig.~\ref{fig:coveragecontours}). Motivated by this, two types of
digital representation or quantization of the protection regions will
be considered: (i) circular protection regions with quantization of
radius; and (ii) non-circular protection regions with super-set or
envelope approximations. These approximations are illustrated in
Fig.~\ref{fig:naivesmart}. The circular protection region model
``marks'' some portion of spatial TV white space as protected region,
and it is ``wasteful'' compared to the non-circular protection region
model. 

In both circular and non-circular models, the protection region has to
be quantized or represented to a larger or envelope-style region to
protect the incumbent. That is, TV white space can be labeled as
protected (due to quantization) but protection region must not be
labeled as unprotected (due to quantization). For this important
regulatory reason, a traditionally well-studied mean-squared error
optimal quantization method \textit{cannot be
employed}~\cite{gershogray}.

\begin{figure}[!htb]
\centering
\includegraphics[scale=0.33]{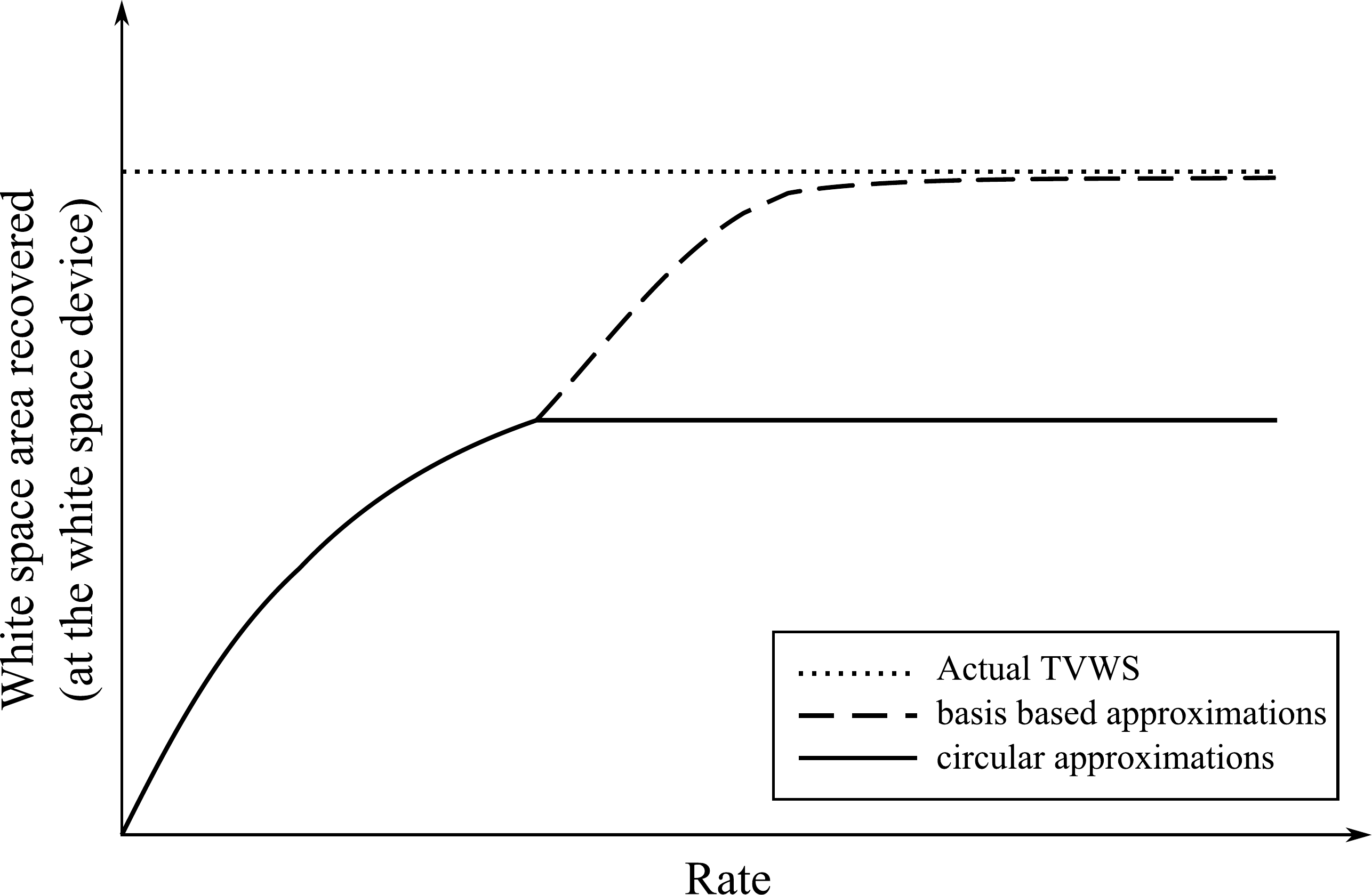}
\caption{\label{fig:overview} The graph qualitatively illustrates the
methods described in this paper. For circular approximations, the
white space area recovered by bounding the protection region with
circles has a gap when compared to the actual white space area. This gap
can be reduced by basis based approximation for the protection
region.}
\end{figure}

\textit{Key contributions:} For the circular model, efficient
quantizer design algorithm will be discussed in this work to ensure
primary's protection even after quantization! By efficient, it is
meant that for a given quantizer precision, the TV white space area
mislabeled (lost) as protection region is minimum.  For the
non-circular model, a basis (such as Fourier) based approach is
detailed to obtain an envelope-approximation to the non-circular
protection region. As explained in Fig.~\ref{fig:naivesmart}, a basis
based approximation will minimize or reduce the loss in TV white space
area beyond the circular (radius based) approximation. The main goal
of our developed techniques is to obtain the qualitative graph
illustrated in Fig.~\ref{fig:overview}. There is a fundamental limit
of how much TV white space area is present. With larger transmission
rate, the database can facilitate a more accurate recovery of the TV
white space region or the protection region. While using the circular
approximation, the recovery of TV white space region will have a gap
from the actual TV white space area.

\begin{figure}[!htb]
\centering
\includegraphics[scale=0.25]{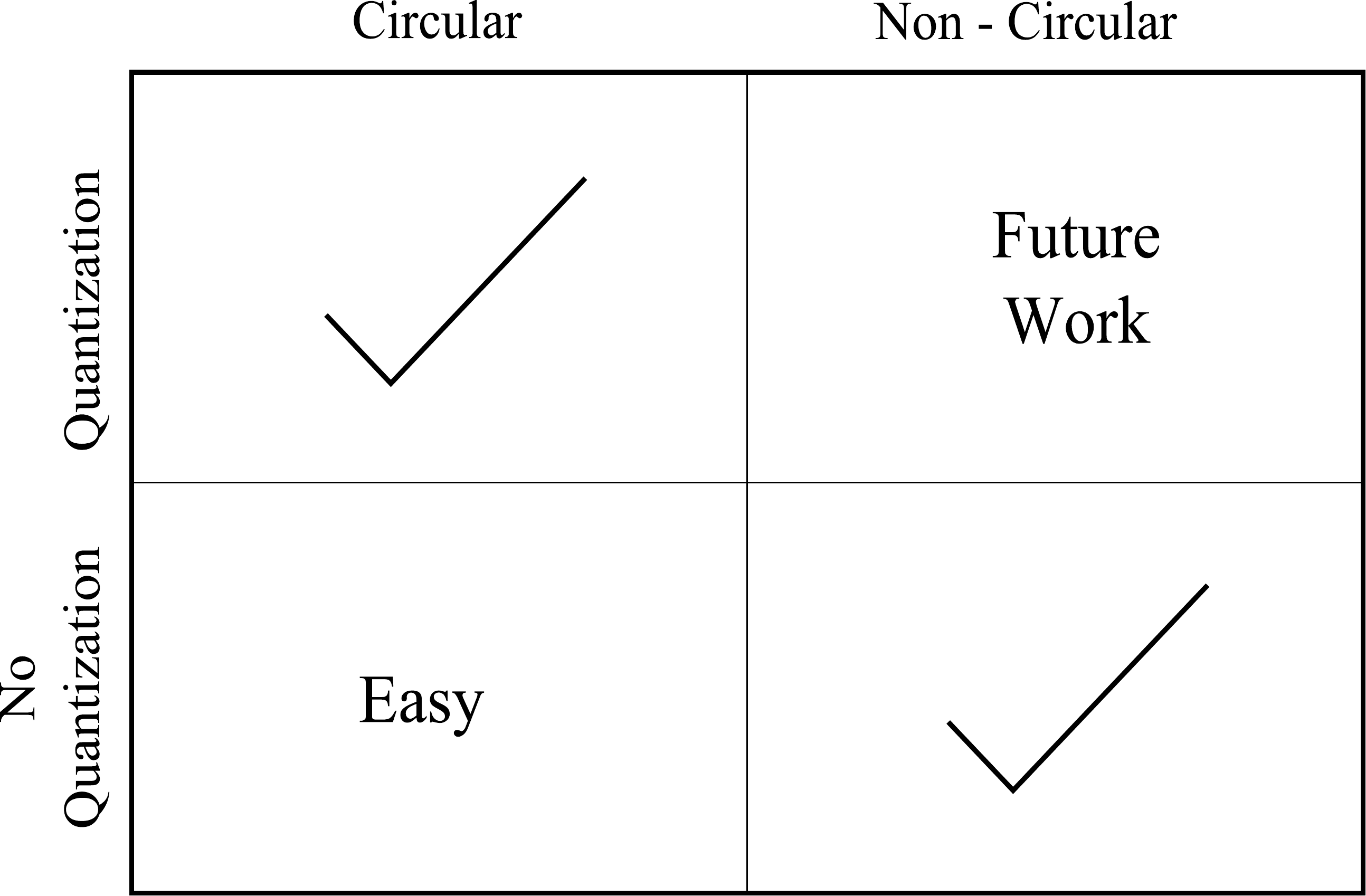}
\caption{\label{fig:problemsolved} The described problem statement in
this paper can be divided into four segments as shown in the figure.
We have addressed the issues of quantization of circular protection
regions and describing the shape of the non-circular regions using a
basis function (without quantization of the coefficients corresponding
to the basis functions).}
\end{figure}

To clarify further, the chart shown in Fig.~\ref{fig:problemsolved}
will be useful to identify our contributions.

\textit{Related work:} Geolocation database are well known in the
literature~\cite{FCC08260S2008,OfcomD2009,gurneyBEKGG2008}. Circular
protection regions for TV white space regions are well known
(see~\cite{harrisonMSH2010}). Primary service contours are available
as databases for countries such as US and
UK~\cite{FCC08260S2008,OfcomD2009}. 

To the best of our knowledge, a broadcast style geolocation database
has not been studied in the literature.  Quantization of real values
where signal is always overestimated as well as envelope style
approximations are also not known in the literature to the best of our
knowledge.

\textit{Organization:} Section~\ref{sec:cost} formulates and solves
the quantization of circular protection regions. Optimal algorithm is
presented  to quantize the protection region, while ensuring
protection for the primary. Section~\ref{sec:noncircular} presents the
basis based envelope-approximation for non-circular protection
regions. Finally, conclusions are presented in Section~\ref{sec:con}.

\textit{Assumptions:} The quantization of transmitters' center is not
considered. In the basis approximation case, this just corresponds to
one extra coefficient being sent and does not affect our main results.
In the circular approximation case, having quantized radius makes the
problem more difficult.  Nevertheless, since TV transmitters are at
fixed locations, this data can be exchanged first before repeated
broadcast of TV white space takes place. It is assumed that the
secondary devices will coexist by some media access control mechanism,
since the broadcast database facilitates only `one-way' communication.
The broadcast database will not register the secondary devices.
Distributed coexistence techniques such as collision sense multiple
access can be used by the secondary devices.

\section{Circular protection regions and their optimal quantization}
\label{sec:cost}

This section deals with the quantization of (circular) protection region
radius. The quantization has to be designed to minimize the white space area
lost due to quantization across all transmitters. It is assumed that the centers
of these protection regions are already available at the receiver, so that only
radius of protection regions have to be quantized and communicated. In case
the protection region is not circular, the na{\"i}ve radius based approximation
scheme depicted in Fig.~\ref{fig:naivesmart} can be used to obtain a radius. 

Let $\RSet := \{r_1, r_2, \ldots, r_n\}$ be the radius of protection
regions. For simple exposition in this paper, it is assumed that this
set $\RSet$ is fixed.  These radius have to be quantized in the set
$\QSet = \{q_1, q_2, \ldots, q_m\}$, where $\log_2 m$ will be the
number of bits being spent to communicate each circular region. These
$\log_2 m$ bits will index various quantization levels in the set
$\QSet$.  Without loss of generality, it is assumed that $q_1, q_2,
\ldots, q_m$ and $r_1, r_2, \ldots, r_n$ are both in an increasing
order.  The quantized radius set $\QSet$ is known to the
broadcast-database transmitter as well as all the secondary white
space devices. The protection region radius set $\RSet$, on the other
hand, is only known to the transmitter.

To \textit{ensure protection} for the primary, $Q(r_i) \geq r_i$ for all
protection radius $r_i \in \RSet$. Unlike in traditional mean-squared or minimax
optimal quantization, where $Q(r)$ is mapped to the closest quantization
level~\cite{gershogray}, quantization level for protection region's radius is
always the \textit{nearest larger level}. Thus, the quantizer design for
protection region radius is \textit{different} than traditional mean-squared
optimal quantizers.

\begin{figure}[!htb]
\centering
\includegraphics[scale=0.33]{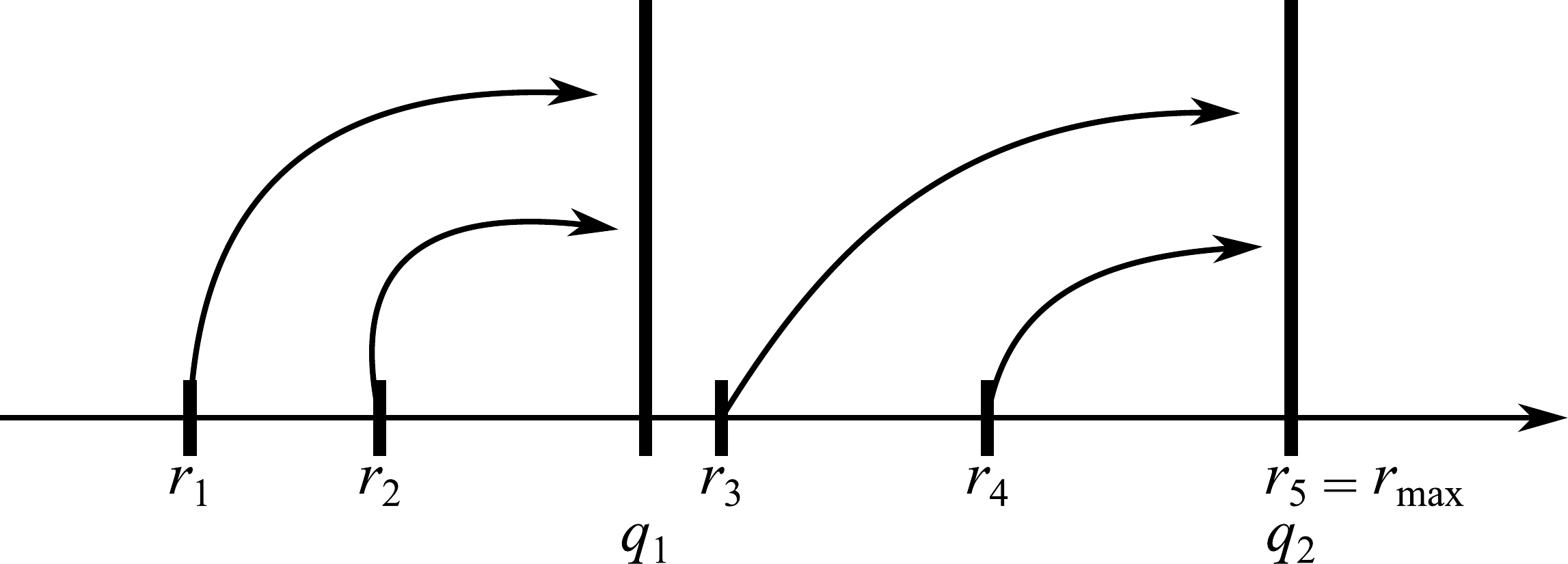}
\caption{\label{fig:radiusPoints} Radius $r_1$, $r_2$ get translated to the
quantization level $q_1$ and $r_3$, $r_4$ get translated to the quantization
level $q_4$ as shown in the image. This is for the purpose of protecting the
primary transmitting device from any harmful interference.}
\end{figure}

The actual area of a circular protection region with radius $r$ is $\pi r^2$.
After quantization, the circular protection region will have a radius of $\pi
Q(r)^2$.  Since $Q(r) \geq r$ by design requirements, so a part of TV white
space region will be \textit{lost} or mislabeled as protection region.  This
motivates the following cost function
\begin{align}
C(\RSet, \QSet) = \pi \sum_{r \in \RSet} \left\{ Q^2(r) - r^2 \right\}
\label{eq:cost}
\end{align}
which signifies the white space area lost due to quantization.  This is
explained with an example having four protection radius and two quantization
levels in Fig.~\ref{fig:radiusPoints}. The radius $r_1, r_2$ get mapped to
$q_1$. Even though $q_1$ is nearer to $r_3$, still $r_3$ gets mapped
to $q_2$. For any given number of quantization levels
$m$, an \textit{optimal} quantization map $Q: \RSet \rightarrow \QSet$ has to be
designed to minimize the lost TV white space area, subject to a primary
protection condition---the actual (unquantized) protection region should be a
\textit{subset} of the quantized protection region.

The cost function in \eqref{eq:cost} can be rewritten as 
\begin{align}
C(\RSet, \QSet) = \pi \left\{ \sum_{i=1}^n Q^2(r_i) - r_i^2 \right\}
\label{eq:loss}
\end{align}
where $Q(r_i)$ maps $r_i$ to the next largest quantization level in the set
$\QSet = \{q_1, q_2, \ldots, q_m\}$. The levels in $\QSet$ have to be chosen to
minimize the lost TV white space area to quantization, or $C(\RSet, \QSet)$ in
\eqref{eq:loss}. The usual technique for minimization will be to evaluate 
\begin{align}
\frac{\partial C(\RSet, \QSet) }{\partial q_j} = 0 \text{ and } \frac{\partial^2
C(\RSet, \QSet)}{\partial q_j^2} > 0
\end{align}
for all $q_j \in \QSet$. Let $r_{\max} = r_n$ be the radius of the largest
protection region in $\RSet$. Then $q_m = r_{\max}$. However, the derivative
does not exist since the set $\RSet$ is discrete and therefore a change in $q_j
\geq r_i$ to $q_j < r_i$ causes a discontinuous change in $C(\RSet, \QSet)$. To
gain further insights, we assume that $\RSet$ is obtained based on i.i.d.~trials
from a (continuous) probability distribution $p(r)$. Then the expected value of
$C(\RSet, \QSet)$ will be minimized. Similar to \eqref{eq:loss}, this expected
cost function is given by
\begin{align}
C(p(r), \QSet) = \pi \int_{r_{\min}}^{r_{\max}} \left\{ Q^2(r) -
r^2\right\} p(r) \mbox{d}r \label{eq:losscont}
\end{align}
where $r_{\min}$ and $r_{\max}$ are the minimum and maximum protection region
radius (over which $p(r)$ is non-zero). Using the quantization levels in
$\QSet$, the cost function in \eqref{eq:losscont} can be rewritten as 
\begin{align}
C(p(r), \QSet) = \pi \int_{r_{\min}}^{q_1} & \left\{ q_1^2 - r^2\right\}
p(r) \mbox{d}r + \ldots + \nonumber \\%
& \pi \int_{q_{m-1}}^{q_m} \left\{ q_m^2 - r^2\right\} p(r)
\mbox{d}r \label{eq:losscont1}
\end{align}
From \eqref{eq:losscont1}, upon taking derivative with respect to $q_j$ in
accordance with the Leibniz integral rule~\cite{rudinp1976}, we get
\begin{align}
\frac {\partial C(p(r), \QSet)}{\partial q_j} = 2q_j \int_{q_{j-1}}^{q_j} p(r)
\mbox{d}r - p(q_j)[q_{j+1}^2-q_j^2 ] \label{eq:derivative}
\end{align}
Recall that $\QSet$ is assumed to be in increasing order, that is, $q_1 < q_2 <
\ldots < q_m$. In the above equation $q_{j-1} \leq q_j \leq q_{j+1}$. If $q_j =
q_{j-1}$, then 
\begin{align}
\frac {\partial C(p(r), \QSet)}{\partial q_j} =  - p(q_j)[q_{j+1}^2-q_j^2 ] < 0
\end{align}
while if $q_j = q_{j+1}$, then 
\begin{align}
\frac {\partial C(p(r), \QSet)}{\partial q_j} = 2q_{j+1}
\int_{q_{j-1}}^{q_{j+1}} p(r) \mbox{d}r > 0.
\end{align}
Using the Intermediate value theorem, with the assumption that $p(r)$ is
continuous so that $\frac {\partial C(p(r), \QSet)}{\partial q_j}$ in
\eqref{eq:derivative} is continuous, it can be deduced that there is some value
$q \in (q_{j-1},q_{j+1})$ such that \begin{align}
\left. \frac {\partial C(p(r), \QSet)}{\partial q_j}\right|_{q_j = q} = 0.
\end{align}
With the assumption that $p(r)$ is continuous, an extrema of $C(p(r), \QSet)$
with respect to $q_j$ lies between $q_{j-1}$ and $q_{j+1}$. To argue that the
extrema is a point of minimum, the Hessian in \eqref{eq:losscont} should be
positive definite, which requires more assumptions on $p(r)$. At this point it
isn't clear what assumptions should be made on $p(r)$, so an empirical approach
is taken. An iterative algorithm will be obtained to find the set $\QSet$ which
minimizes $C(r, \QSet)$ or $C(\RSet, \QSet)$. This iterative algorithm is
explained next for minimizing $C(\RSet, \QSet)$, i.e., assuming that $\RSet$ is
given. 

At first, note that the largest quantization level must be equal to
the largest protection radius, i.e.,
\begin{align}
q_m  = r_n \label{eq:qm}
\end{align}
since the largest values of $Q(r_i)$ need not exceed $r_n$ (the
maximum protection radius) in \eqref{eq:loss}. Apart from $q_m$, each
$q_{j}$ will be chosen from the discrete-set $\RSet$ only. This can be
understood using Fig.~\ref{fig:radiusPoints}. If $q_1$ is chosen
between $r_2$ and $r_3$, then the cost contribution of $q_1$ is $q_1^2
- r_1^2 + q_1^2 - r_2^2$, which gets minimized when $q_1 = r_2$.

At a high-level, one notes that $q_1$ has $(n-m)$ choices in $\RSet$,
subsequently $q_2$ has $(n-m-1)$ choices in $\RSet$, and so on. So the
total number of choices for the entire set $\QSet$ is $(n-m) (n - m
-1) \ldots (1) = (n-m)!$, which is quite large. A fast algorithm will
be developed next to solve the selection of $\QSet$.

From \eqref{eq:derivative}, for minimum (or an extrema) point the following
equation should be satisfied
\begin{align}
2q_j \int_{q_{j-1}}^{q_j} p(r) \mbox{d}r - p(q_j)[q_{j+1}^2-q_j^2 ] = 0.
\label{eq:evenoddbasic}
\end{align}
From the above, it is noted that  $q_j$ depends only on $q_{j-1}$ and $q_{j+1}$.
That is, if the odd elements in the set $\QSet$ are fixed, then the even
elements can be found separately. Similarly, if the even elements in the set
$\QSet$ are fixed, then the odd elements can be found separately. This results
in a separable optimization algorithm as detailed next. The counterpart of
\eqref{eq:evenoddbasic} for the discrete case will be highlighted first. The
cost function in \eqref{eq:loss} can be rewritten as
\begin{align}
\frac{1}{\pi} C(& \RSet, \QSet) \nonumber \\
& = \sum_{r\in \RSet} Q^2(r) - \sum_{r\in \RSet} r^2  \nonumber \\
& = \sum_{r \in (q_{j-1}, q_j]} Q^2(r) + \sum_{r \in (q_{j}, q_{j+1}]} Q^2(r) +
E_j - \sum_{r\in \RSet} r^2 \nonumber \\
& = \sum_{r \in (q_{j-1}, q_j]} q_j^2 + \sum_{r \in (q_{j}, q_{j+1}]} q_{j+1}^2
+ E_j - \sum_{r\in \RSet} r^2 \label{eq:seploss} 
\end{align}
where the term $E_j$ is positive, depends on $q_1, \ldots, q_{j-1}, q_{j+1},
\ldots, q_m$, and is independent of $q_j$. In \eqref{eq:seploss}, it is also
noted that  $r \in (q_{j-1}, q_j]$ will get quantized to $q_j$ and $r \in
(q_{j}, q_{j+1}] $ will get quantized to $q_{j+1}$.  Let the number of
protection region radius $ r \in (q_{j-1}, q_{j+1}]$ be $n_j$ and number of
radius $ r \in (q_{j-1}, q_{j}]$ be $k_j$.
Then \eqref{eq:seploss} can be rewritten as
\begin{align}
\frac{1}{\pi} C(\RSet, \QSet) & = k_j q_j^2 + (n_j - k_j) q_{j+1}^2 + E_j - \sum_{r\in
\RSet} r^2 \label{eq:kjnj}
\end{align}
The last term is independent of $\QSet$, while the second last term $E_j$ is
independent of $q_j$ and can be ignored during optimization. Since $q_{j-1}$ and
$q_{j+1}$ are fixed, so is $n_j$. Therefore, the only choice variables are
$q_{j}$, which subsequently determines $k_j$ as well. The minimization of $k_j
q_j^2 + (n_j - k_j) q_{j+1}^2$ and subsequently the expression in \eqref{eq:kjnj} (for
fixed $q_{j-1}$ and $q_{j+1}$) can be performed by an exhaustive search over
various values of $q_j$ in between $q_{j-1}$ and $q_{j+1}$. In summary, for
given fixed values of $q_{j-1}$ and $q_{j+1}$,  the value of $q_j$ that
(locally) minimizes $C(\RSet, \QSet)$ can be found out by an exhaustive search.

This motivates the following \textit{Even-Odd} algorithm for the
minimization of cost function in \eqref{eq:loss}, subject to the
condition that quantized protection radius is always larger than the
actual protection radius:
\begin{enumerate}
\item A random initialization for the quantization levels in the set
$\QSet$ is assumed. It must be noted that the quantization levels
belong in the set $\RSet$.
\item The largest quantization level $q_m$ is fixed to the largest
protection radius $r_n$.
\item After a random initialization, the even quantization levels are
fixed and the odd quantization levels are exhaustively searched
according to the process outlined in \eqref{eq:kjnj}. It is restated
that the exhaustive search for each quantization level is separate.
This results in optimal values for odd quantization levels with
respect to the cost function in \eqref{eq:loss}. 
\item The (locally) optimized values of the odd quantization levels,
as obtained in the previous step, are fixed. The even quantization
levels are now exhaustively searched according to the process outlined
in \eqref{eq:kjnj}. This results in optimal values for the even
quantization levels with respect to the \eqref{eq:loss}.
\item The steps 3 and 4 above are used in an iterative manner, 
till the quantization levels do not change.  The resultant
quantization levels minimize the desired cost function stated in
\eqref{eq:loss}.
\end{enumerate}
The simulation results are presented next. The data used in our
simulation experiments is outlined first.

\textit{Data used for experiments:} For circular protection regions,
two sets of data were available to us.  The first data-set contains
the protected service contours' bounding radius calculated using
protection and pollution viewpoints~\cite{harrisonMSH2010}. They make
use of the latitude, longitude, EIRP and HAAT to calculate this radius.
This set of radius is available for all the channels between $2$ to
$51$ ($49$~channels) of United States, where transmission by a white
space device is permitted by the FCC. There were $8141$ protection
regions, and hence radius, in total.  The second data-set contains the
protected service contours' bounding radius obtained for
India~\cite{naikSKKQ2014}. There are no white space regulations in
India as of now. The protection radius are available only for $15$
channels in the UHF Band-III ($470$-$590$MHz).  There were $374$
protection regions, and hence radius, in total for India.

These datasets are used to analyse the recovered TV white space area
and test our optimal quantization Even-Odd algorithm.  As India has
extensive rural areas with negligible broadband services, Internet
connection is extremely unreliable. It is once again emphasized that a
broadcast based TV white space geolocation database will be quite
useful for such scenarios. 

If $b$ bits are used to index each radius in the set $\RSet$, then $m
= 2^b$. It is recalled that the radius set $\RSet$ is only known to
the database, points in $\RSet$ will be mapped into $m$ quantization
levels, and $\QSet$ is agreed upon between the broadcasting
geolocation database and the secondary devices. For comparison of our
Even-Odd algorithm discussed in Section~\ref{sec:cost}, a uniform
scalar quantizer is used. For uniform scalar quantizer, the
quantization levels are $\QSet = \{r_{\min} + \Delta, r_{\min} + 2
\Delta, \ldots, r_{\max}\}$ where 
\begin{align}
\Delta = \frac{r_{\max} - r_{\min}}{m}.
\end{align}
All the values of radius ($r$) greater than
${i-1}^{\mbox{\footnotesize th}}$ level and less than or equal to
$i^{\mbox{\footnotesize th}}$ level are translated to the
$i^{\mbox{\footnotesize th}}$ level in order to protect the primary.

\begin{figure}[!htb]
\centering
\includegraphics[scale=0.45]{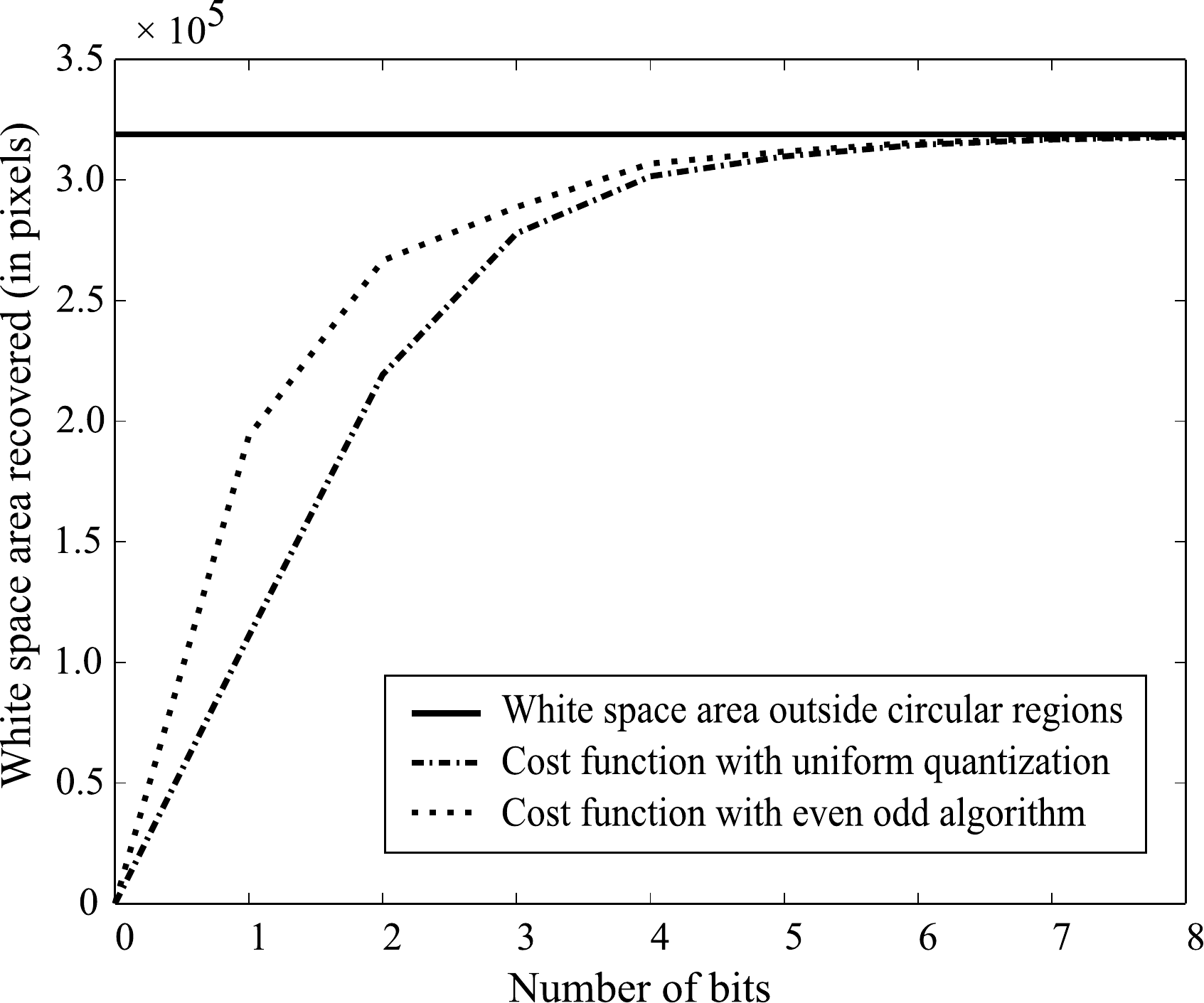}
\caption{\label{fig:UniformVsEO} The recovered white space area for
various values of bits/protection region is plotted for the data-set
from United States~\cite{harrisonMSH2010}. The solid line represents
the actual real valued white space area. For smaller values of
bits/protection region, the Even-Odd algorithm based quantizer
outperforms the uniform quantizer.}
\end{figure}

\begin{figure}[!htb]
\centering
\includegraphics[scale=0.45]{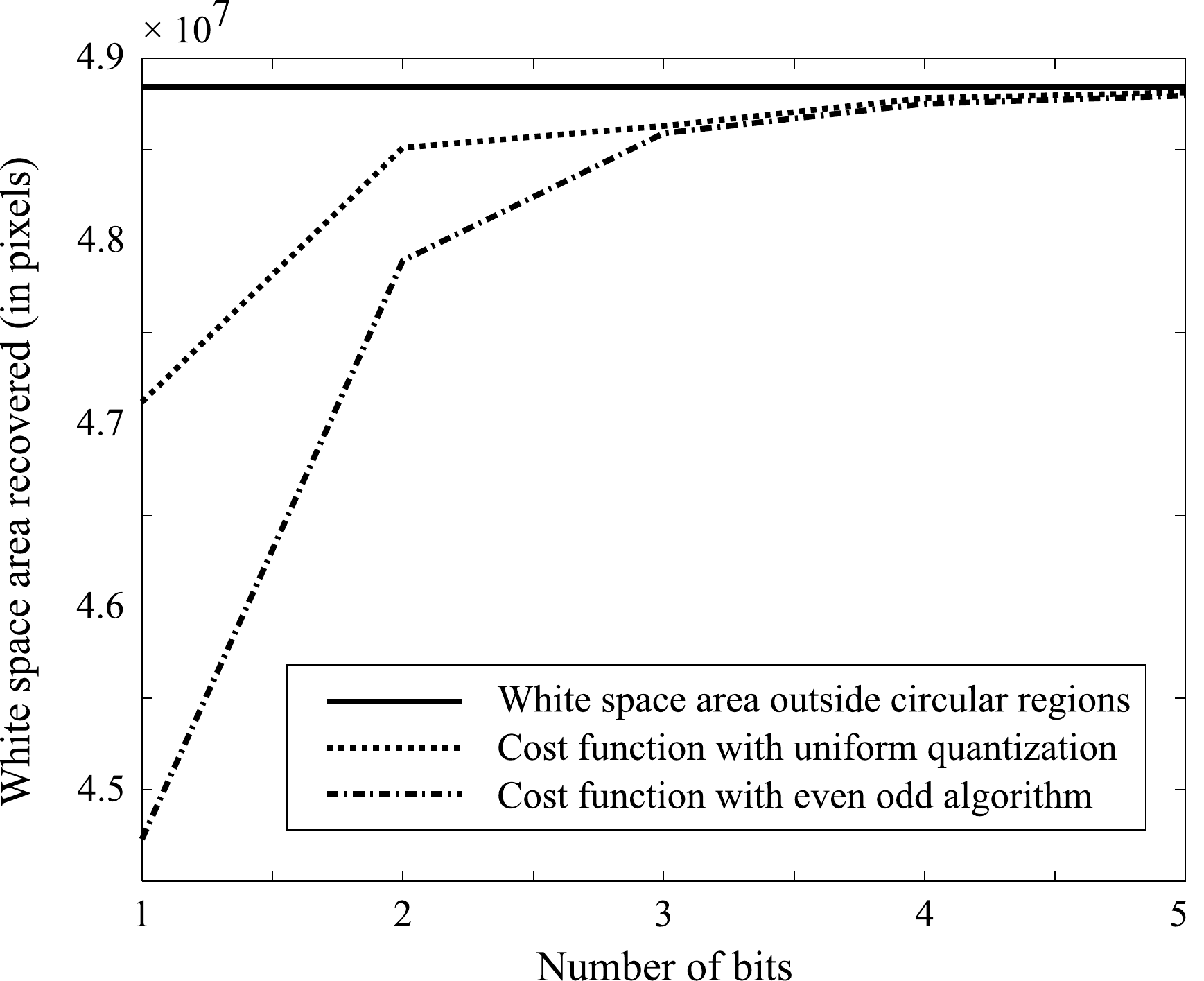}
\caption{\label{fig:IndiaPlot} The recovered white space area for
various values of bits/protection region is plotted for the data-set
from India~\cite{naikSKKQ2014}. The solid line represents the actual
real valued white space area. As expected, the Even-Odd algorithm
based quantizer outperforms the uniform quantizer.}
\end{figure}

The results obtained by applying our algorithm on the data-set from
United States are illustrated in Fig.~\ref{fig:UniformVsEO}.
Increasing the number of bits decreases the area of lost white space
region to quantization, and with $5$ bits (per protection radius) or
$32$ quantization levels most of the white space area can be
recovered. Since there are about $8141$ protection radius, they can be
communicated in a lossless manner with $14$ bits per protection
radius.  The results obtained by applying our algorithm on the
data-set from India are illustrated in Fig.~\ref{fig:IndiaPlot}.  With
$4$ bits (per protection radius) or $16$ quantization levels recover
most of the white space area. Since there are about $373$ protection
radius, they can be communicated in a lossless manner with $8$-$9$
bits per protection radius.

\begin{figure}[!htb]
\centering
\includegraphics[scale=0.45]{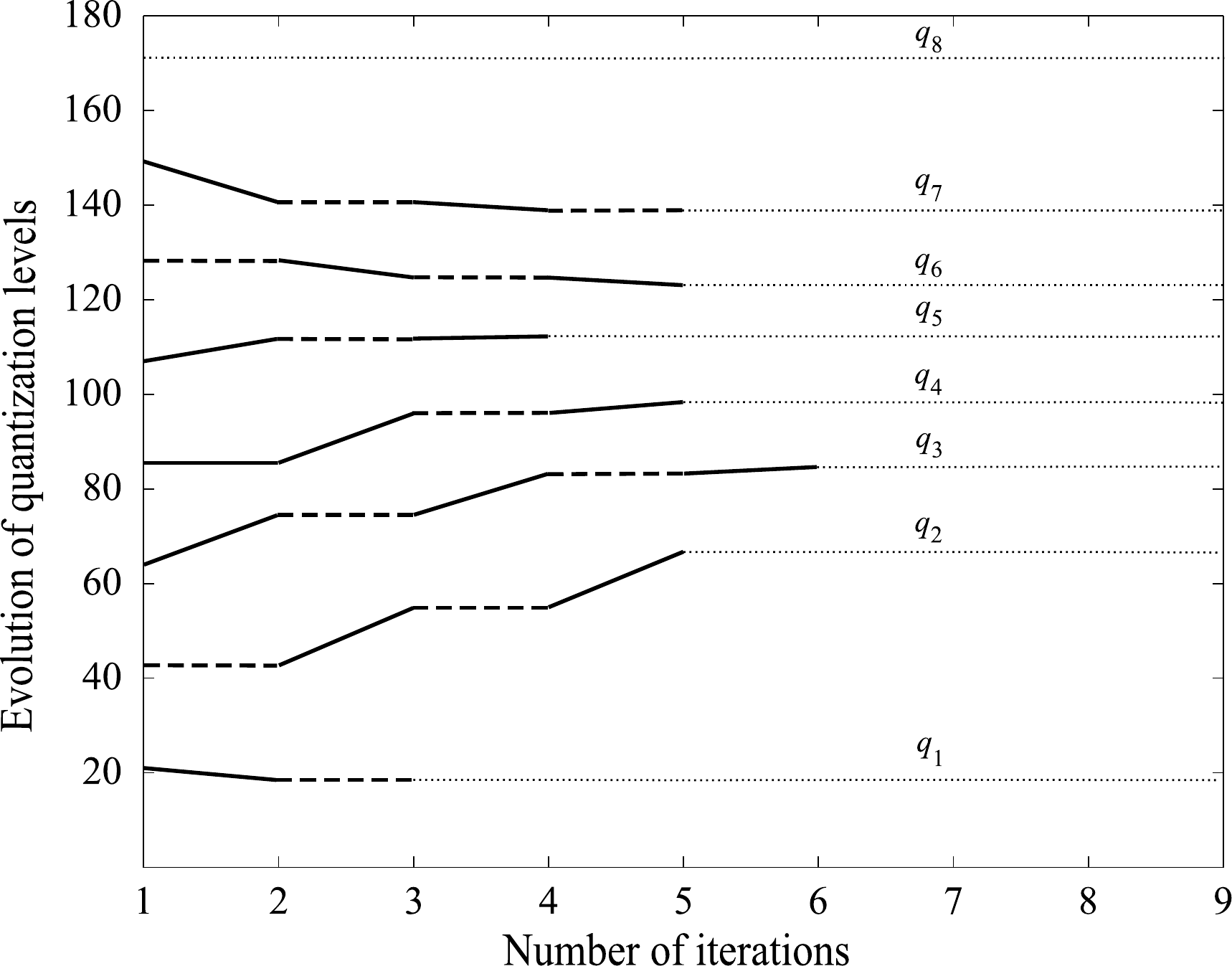}
\caption{\label{fig:QuantizationLevels} The evolution of quantization
levels in the Even-Odd algorithm is illustrated. The initial points
are obtained from a uniform quantizer. The largest level $q_8$ is
always equal to $r_{\max}$. In the odd steps $(q_1, q_3, q_5, q_7)$
are calculated while $(q_2, q_4, q_6)$ are held fixed. The roles are
reversed in the odd steps. The movement is illustrated by solid line,
the fixed behavior is illustrated by a dashed line, and convergence of
the quantization level is shown with a dotted line.}
\end{figure}

The evolution of quantization levels in our Even-Odd algorithm is
illustrated in Fig.~\ref{fig:QuantizationLevels}. The initial
quantization levels are obtained by using a scalar quantizer for $b =
3$. The quantization levels obtained using the algorithm, help in
recovering more white space area as compared to uniform quantization
for every bit that is sent as illustrated in
Fig.~\ref{fig:UniformVsEO} and Fig.~\ref{fig:IndiaPlot}.

\section{Envelope approximation for non-circular protection regions}
\label{sec:noncircular}

For some non-circular protection regions,  its circular approximation
area can be as large as $30$\% when compared to the protection
region's area.  This area can be scavenged by the secondary if a
better approximation method is used. In this section, protection
region representation will be considered beyond the circular
approximation explained in the Introduction. 

\begin{figure}[!htb]
\centering
\includegraphics[scale=1.0]{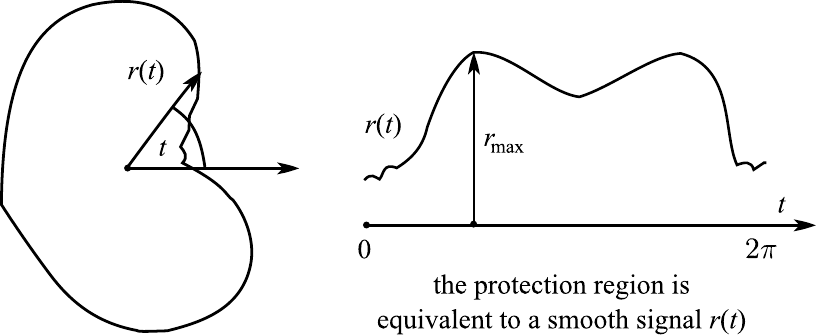}
\caption{\label{fig:better_approx} A non-circular protection region can
be viewed as a one-dimensional waveform $r(t)$ with respect to the
angle $t$ as shown.}
\end{figure}

As explained in Fig.~\ref{fig:problemsolved}, for non-circular regions
quantization is not considered in this work. This section will focus
on obtaining basis-based envelope approximation as depicted in
Fig.~\ref{fig:naivesmart}. To the best of our knowledge, such
approximations are not present in the quantization literature.
Consider a non-circular but smooth shape as depicted in
Fig.~\ref{fig:better_approx}. Let its centroid be at the origin. Then,
the shape can be parametrized by a waveform $r(t)$ as shown in the
figure. With the knowledge of origin (the center), the waveform $r(t)$
is equivalent to the non-circular protection region. The waveform
$r(t)$ is periodic (or has finite support) and it is expected to be
smooth; so, it can be represented via any basis function suitable for
smooth signals on a compact support~\cite{mallatSA2009}. That is,
\begin{align}
r(t) = \sum_{ k = - \infty}^{\infty} a_k \phi_k(t) \label{eq:basis}
\end{align}
where $\phi_k(t), k \in \zZ$ form the set of basis functions suitable
for representing smooth waveforms with support in $[0,2\pi]$. The set
$\zZ$ denotes the set of integers. Fourier series happens to be one
choice of basis functions, which will be used in this current work for
simplicity of exposition.  For Fourier basis, the expansion in
\eqref{eq:basis} reduces to 
\begin{align}
r(t) = \sum_{ k = - \infty}^{\infty} a_k \exp{ (j k t)}
\label{eq:fourier}
\end{align}
The Fourier series coefficients are given by 
\begin{align}
a_k = \frac{1}{2\pi}\int_{0}^{2\pi} r(t) \exp{(-j k t)} \mbox{d} t
\end{align}
Since $r(t)$ is real-valued the coefficients $a_k$ and $a_{-k}$ will
be related by conjugate symmetry. That is $a_k = \bar{a}_{-k}$. So,
only $a_k, k \geq 0$ needs to be communicated. 

Our main innovation in this section is an envelope approximation for
the waveform $r(t)$. In other words, $r(t)$ has to be approximated to
$r_{\approx}(t)$ such that $r_{\approx}(t)$ is always larger than
$r(t)$. To this end, the following approach is used. Let $r_K(t)$ be
the $K+1$ coefficient based Fourier series approximation, that is,
\begin{align}
r_K(t) = \sum_{ k = - K}^{K} a_k \exp{(jkt)}. \label{eq:Kapprox}
\end{align}
Define $e_K$ as 
\begin{align}
e_K = \max_{0 \leq t \leq 2\pi} r(t) - r_K(t)
\end{align}
It is noted that $e_K$ is the maximum error between $r(t)$ and
$r_K(t)$. So if
\begin{align}
r_{\approx}(t) = r_K(t) + e_K \label{eq:rapprox}
\end{align}
then $r_{\approx}(t) \geq r(t)$, and $r_{\approx}(t)$ has $2K+1$
Fourier coefficients given by $a_{-K}, \ldots, a_{-1}, e_K + a_0, a_1,
\ldots, a_K$. Of these, due to conjugate symmetry, only $K+1$
coefficients need to be sent. The coefficients that need to be sent are:
$[a_0 + e_K, a_1, a_2, ..., a_K]$ to communicate an envelope approximation
$r_{\approx}(t)$ for $r(t)$. Experimental evaluation with this scheme
is explained next. 

\begin{figure*}[!htb]
\centering
\includegraphics[scale=0.43]{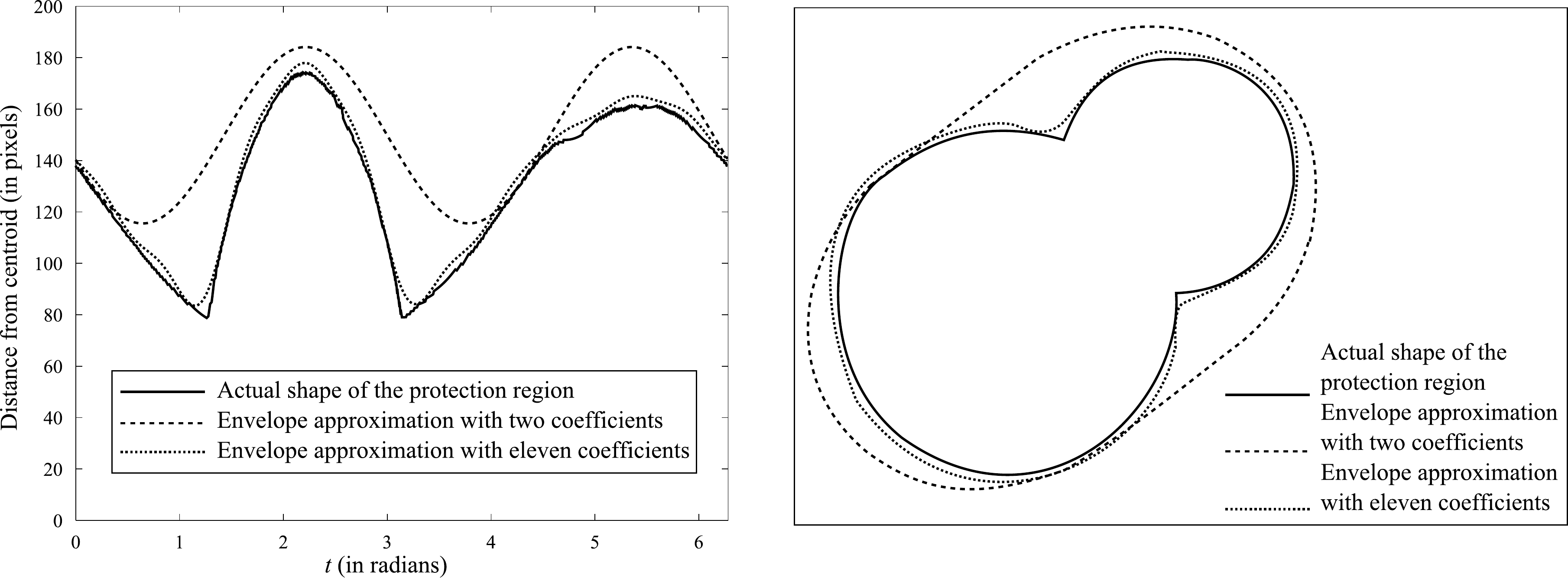}
\caption{\label{fig:radiusApprox} The approximation in
\eqref{eq:rapprox} is illustrated. Consider the shape given by solid
line in the right graph. The $r(t)$ corresponding to it is given by
the solid line in the left graph. Using $K = 2$ and $K = 11$, two
different $r_{\approx}(t)$ are plotted. Observe that for both values
of $K$, the approximation is larger than $r(t)$ for each value of $t$.
This property is expected due to design of the approximation in
\eqref{eq:rapprox}. This property also ensures that the approximate
protection region includes (or, is a superset of) the actual
protection region.}
\end{figure*}
There is no convenient data-set available for the shapes $r(t)$ for
the $8141$ protected regions in the United States. To test the Fourier
series based envelope approximation method, we created a small
data-set for Channel~$2$ using the TV white spaces US Interactive Map
of Spectrum Bridge~\cite{spectrumWEB}. Across United States, there are $57$ protected
service contours (excluding Alaska). These contours were hand-picked
into images and the coverage region was segmented (using image
processing techniques) to obtain various shapes $r(t)$ for these $57$
transmitters in Channel~$2$. Then, the Fourier basis based envelope
approximation technique was applied. An example of Fourier basis based
envelope is shown in Fig.~\ref{fig:radiusApprox}. Observe that
$r_{\approx}(t)$ for various values of $K$ always includes the actual
protection region as a subset.

\begin{figure}[!htb]
\centering
\includegraphics[scale=0.45]{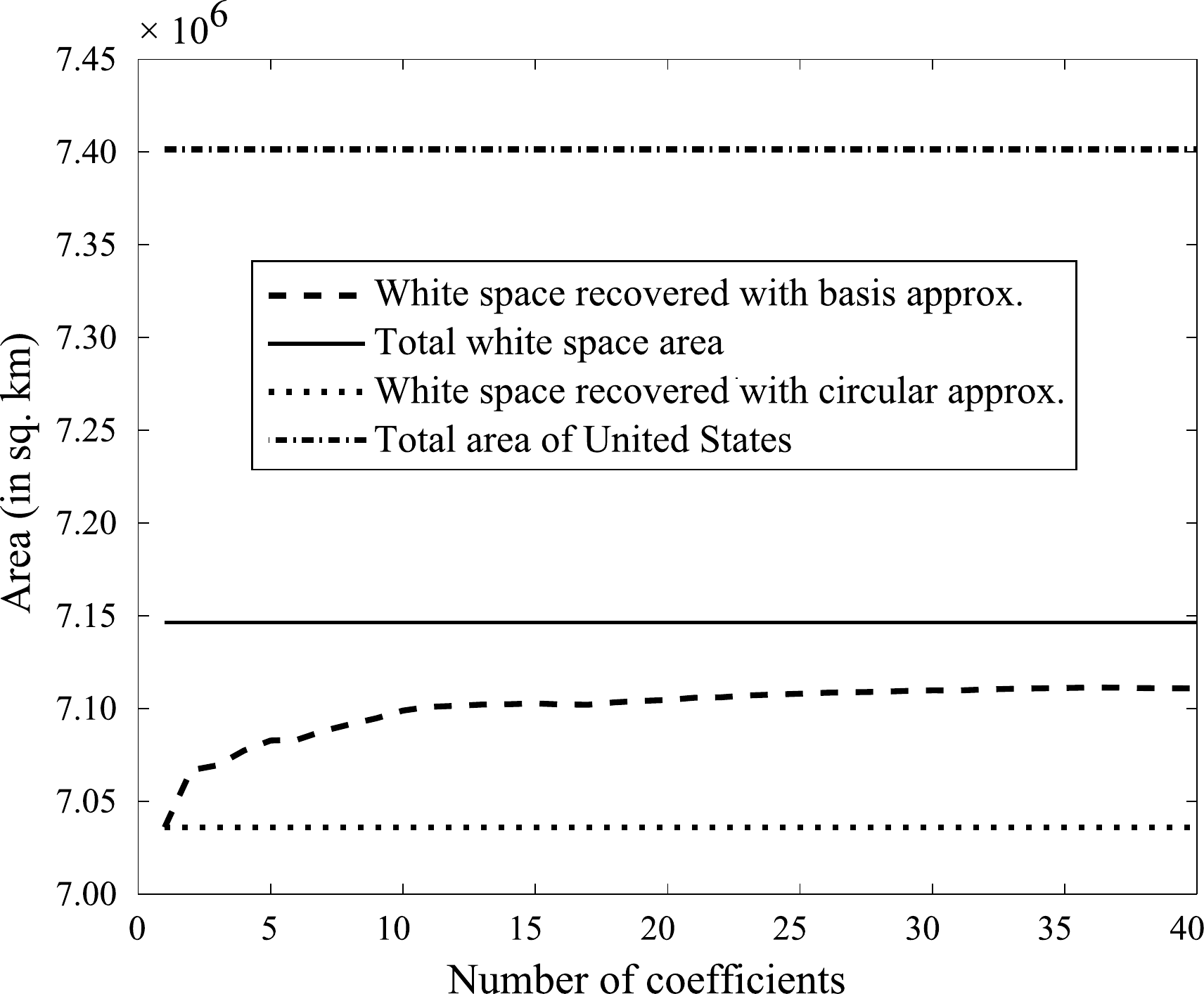}
\caption{\label{fig:Fourier} This graph is applicable for $57$
protection regions in Channel~2 in United States (except Alaska).  The
dotted line is the white space area that can be obtained using
circular approximations, which is less than the actual white space
area (in solid line). If more Fourier series coefficients for each
protection region is sent, then the extra white space recovered above
the circular approximation is illustrated.}
\end{figure}

In the absence of real-valued data (images obtained have pixels!), a
Riemann approximation~\cite{rudinp1976} is used to calculate the
Fourier series coefficients as well as the protection region's area of
the extracted service contours.  As there is a limitation on the
resolution of the images that can be obtained from the Interactive
Map, we did observe some issues due to pixelization (see
Fig.~\ref{fig:Fourier}). With a higher-resolution data-set, we will
obtain better results. But such data-set is not available. Due to
pixelization, in Fig.~\ref{fig:Fourier}, the dashed line signifying
white space recovered with basis approximation does not converges to
the solid line depicting total white space area. Based on the data-set
we could scavenge, the each pixel width is 0.412km.

The results obtained by sending Fourier coefficients can be seen in
Fig.~\ref{fig:Fourier}. This graph is for the white space region in
Channel~2 of the TV spectrum in United States (except Alaska).  The
circular approximation scheme sends one coefficient $K = 1$, but there
is a gap between total white space area available and the white space
area that can be recovered by circular approximation. As the number of
coefficients sent is increased, $K = 2$ and higher, the Fourier basis
scheme fills in the gap between circular approximation and actual
white space area. It can be observed that the actual protection region
is $(7.401 - 7.152) \times 10^6 = 2.49 \times 10^5$ sq.~km. Circular
approximation (with high bit-rate quantization) labels $(7.401 - 7.030)
\times 10^6 = 3.71 \times 10^5$ sq.~km.~as protection region. So, the
basis based approximation method reduces the size of protection region
by about $(3.71 - 2.49)/2.49 = 49.0\%$.

\section{Conclusions}
\label{sec:con}

A geolocation database that broadcasts the TV white space or the
primary services protection regions on rate-constrained digital
channel was considered. The key issue addressed was quantization or
digital representation of the protection regions. It was observed that
the protection regions can be circular or non-circular. For circular
protection regions, a fast algorithm for optimal quantizer design was
developed. The algorithm minimizes the white space area identified as
protection region, while ensuring that protection region is not
labeled as white space region due to quantization. For non-circular
protection regions, a basis based approximation was developed. A
procedure for obtaining envelope-type approximation for protection
region was developed, which depends on very few number of basis
coefficients. The approximation methods were tested using three
experimental data-sets. These data-sets included circular protection
regions across all TV channels in the United States, circular
protection regions in the UHF band TV channels in India, and
non-circular protection regions in Channel~2 in the United States.

\section*{Acknowledgments}

The authors benefited tremendously from stimulating discussions with
Prof.~Anant Sahai, EECS, University of California, Berkeley, CA on
this topic.

This research is supported by the Ford Foundation.

\bibliographystyle{IEEEtran}


\end{document}